\documentclass[pra,amssymb,letterpaper,twocolumn,nofootinbib]{revtex4}

\usepackage{amsmath}
\usepackage{epsfig}
\usepackage{amsfonts}
\usepackage{graphicx}

\def\nn{\nonumber}
\def\>{\rangle}
\def\<{\langle}

\def\be{\begin{equation}}
\def\ee{\end{equation}}
\def\bea{\begin{eqnarray}}
\def\eea{\end{eqnarray}}

\begin{document}
\title{A 2 rebit gate universal for quantum computing}

\author{Terry Rudolph}
\email{rudolpht@bell-labs.com}
\author{Lov Grover}
\affiliation{Bell Labs, 600-700 Mountain Ave., Murray Hill, NJ 07974, U.S.A.}

\date{\today}

\begin{abstract}
We show, within the circuit model, how any quantum computation
can be efficiently performed using states with only real
amplitudes (a result known within the Quantum Turing Machine
model). This allows us to identify a 2-qubit (in fact
2-\textit{re}bit) gate which is universal for quantum computing,
although it cannot be used to perform arbitrary unitary
transformations.

\end{abstract}

%\pacs{03.67.Dd}

\maketitle

Performing universal quantum computation is generally equated with
the ability to build up arbitrary unitary transformations acting
on $n$ qubits,  out of a set of unitary transformations that act
on a small number of qubits  at a time. Deutsch \cite{deutsch}
originally presented a single 3 qubit universal quantum gate from
which all
$n$ qubit unitary transformations could be built. It was
subsequently shown that two qubit gates suffice
\cite{divincenzo}.\footnote{While this was surprising from a computer
science perspective (because universal reversible classical
computing provably requires a three bit gate), it would have been
distressing from a physicist's perspective had it been otherwise
- all fundamental physical interactions are of a ``two-body''
form!}

In general, when evaluating a new proposal for implementing
quantum computation, the standard procedure is to check whether
one can perform (i) a controlled-NOT (CNOT) operation between two
qubits, and (ii) arbitrary single qubit unitary transformations.
If so, then universal quantum computing is certainly possible.
Recently some beautiful ideas for implementing quantum
computation by performing measurements on appropriate states have
been presented \cite{raussendorf}. However the general principle
of proving universality in accordance with the ability to obtain
evolution corresponding to (i) and (ii) has still been followed.

The ability to perform a CNOT and arbitrary single qubit gates
allows one to evolve an $n$ qubit state to any point in the
Hilbert space, i.e. it allows the construction of arbitrary
unitary transforms. However it has been shown within the Quantum
Turing Machine model that universal quantum computing can be
performed using only \emph{real} amplitudes
\cite{adleman}.

The Quantum Turing Machine model is not particularly intuitive for
either thinking about construction of a practical quantum
computer, nor for design of quantum algorithms. The purpose of
this note is to point out how any quantum computation can be
simply translated into a quantum circuit in which all quantum
states and gates are real. In particular we will show that the
following two qubit gate (written in the computational basis) is
universal for quantum computing: \be\label{G} G=\begin{pmatrix}
  1 & 0 & 0 & 0 \\
  0 & 1 & 0 & 0 \\
  0 & 0 & \cos\phi & -\sin\phi \\
  0 & 0 & \sin\phi & \cos\phi
\end{pmatrix},
\ee
where $\phi$ is some irrational multiple of $\pi$. Note that this
gate does
\emph{not} allow arbitrary unitary operations to be performed, as
can be easily deduced from the fact that the matrix entries are
real - and thus it can only produce superpositions of states with
real amplitudes. In other words, such a gate cannot in fact
evolve us through much, in fact most, of an $n$ qubit hilbert
space.

In addition to the ability to perform a gate $G$, we need the
standard assumption that the qubits can be prepared initially in
the computational basis. Generally one assumes they can be
prepared in the state $|0\>$, however we will assume here that
they can be prepared in the state $|1\>$. This means that by using
an ancilla in the state $|1\>$ as the control bit, we can
implement the single
qubit gate $\left(\begin{smallmatrix}\cos\phi & -\sin\phi \\
\sin\phi &
\cos\phi\end{smallmatrix}\right)$ by implementing $G$.

The choice of $\phi$ to be an irrational multiple of $\pi$ is well
known to allow us (by a polynomial number of repeated
applications) to efficiently approximate the gate \be\label{F}
F(\theta)=\begin{pmatrix}
  1 & 0 & 0 & 0 \\
  0 & 1 & 0 & 0 \\
  0 & 0 & \cos\theta & -\sin\theta \\
  0 & 0 & \sin\theta & \cos\theta
\end{pmatrix},
\ee for any desired value of $\theta$. Our goal therefore is to
show that a gate of the form $F(\theta)$ suffices for universal
quantum computation.\footnote{In fact the the use of irrational
numbers here is not strictly necessary, following the argument of
\cite{adleman} certain rational numbers will suffice.}

We begin by showing how any quantum computation which uses
complex amplitudes, can be replaced by one which is as efficient
(in the complexity theoretic sense), but which makes use of only
real amplitudes. Imagine the standard quantum algorithm involves
the creation at some point of the state:
\be\label{sf}
|\psi\>=\sum_j r_j e^{i\theta_j}|j\>.
\ee
If we introduce an ancilla 2-level qubit, the orthonormal states
of which we label
$|R\>$ and $|I\>$, then an equivalent state for the purposes of
quantum computing is
\be\label{ef}
|\psi_E\>=\sum_j r_j \cos\theta_j|j\>|R\>+r_j\sin\theta_j|j\>|I\>.
\ee
 The purpose of the two-level ``R-I'' ancilla bit is to keep track of
Real and Imaginary parts of the amplitudes which appear in
(\ref{sf}). We will use the terminology that the state (\ref{ef})
is the \emph{encoded form} of (\ref{sf}). Note that the
probability of obtaining the state $|j\>$, upon a measurement in
the computational basis, is
$r_j^2$ for both
$|\psi\>$ and $|\psi_E\>$, and of course the amplitudes of
$|\psi_E\>$ are real.

We need to show that if an efficient algorithm is implemented such
that our standard quantum computer now undergoes an evolution
\[
|\psi\>\rightarrow|\psi'\>=\sum_j r_j' e^{i\theta_j'}|j\>,
\]
then an efficient set of gates $F(\theta)$ can be found such that
\[
|\psi_E\>\rightarrow|\psi_E'\>=\sum_j r_j'
\cos\theta_j'|j\>|R\>+r_j'\sin\theta_j'|j\>|I\>.
\]

We begin by showing how to map arbitrary single qubit gates from
the standard to the encoded form, using only $F(\theta)$ gates.
In the standard form an arbitrary single qubit rotation can be
performed using some combination of the gates
\[
R_z(\tau)=\begin{pmatrix}
  1 & 0 \\
  0 & e^{\i\tau}
\end{pmatrix},
R_y(\tau)=\begin{pmatrix}
  \cos\tau & -\sin\tau \\
  \sin\tau & \cos\tau
\end{pmatrix}.
\]
Consider the action of $R_z(\tau)$ on a single qubit state in
standard form:
\bea
|\psi\>&=&r_0e^{i\theta_0}|0\>+r_1e^{i\theta_1}|1\>\nn\\
 &{\rightarrow}& r_0e^{i\theta_0}|0\>+r_1e^{i(\theta_1+\tau)}|1\>.\nn\\
\eea
The equivalent evolution in encoded form is
\bea
|\psi_E\>&=&r_0\cos\theta_0|0\>|R\>+r_0\sin\theta_0|0\>|I\>\nn\\
&+&r_1\cos\theta_1|1\>|R\>+r_1\sin\theta_1|1\>|I\>\nn\\
&\rightarrow&r_0\cos\theta_0|0\>|R\>+r_0\sin\theta_0|0\>|I\>\nn\\
&+&r_1\cos(\theta_1+\tau)|1\>|R\>+r_1\sin(\theta_1+\tau)|1\>|I\>.\nn\\
\eea
This evolution can be achieved by performing a 2-qubit gate
$F(\tau)$, where the \emph{target} qubit is the R-I ancilla.
i.e. we control on the first qubit; in the standard encoding this
would be the qubit upon which $R_z(\tau)$ acted.

If we now consider the action of $R_y(\tau)$ on a qubit in
standard form, it is simple to see that exactly the same
evolution can be achieved in encoded form by applying the same
gate
$R_y(\tau)$ to the equivalent qubit. (Use of the R-I ancilla is
not necessary; this gate does not mix real and imaginary parts of
amplitudes.) As mentioned previously, this gate can be
implemented by $F(\tau)$.

Having seen how to map the action of arbitrary single qubit gates
in a standard quantum computation to implementations of
$F(\theta)$ on qubits in the encoded form, it remains to be shown that we
can implement the encoded form of a non-trivial 2-qubit gate.
Generally one chooses a controlled NOT gate, however we will use
here the gate
$F(\pi/2)$ (which combined with arbitrary single qubit rotations is universal for
quantum computing in the standard form). Acting on two qubits in
standard form, the gate $F(\pi/2)$ implements
\bea\label{aa}
|\psi\>&=&r_0e^{i\theta_0}|00\>+r_1e^{i\theta_1}|01\>+r_2e^{i\theta_2}|10\>+r_3e^{i\theta_3}|11\>,\nn\\
    &\rightarrow&r_0e^{i\theta_0}|00\>+r_1e^{i\theta_1}|01\>+r_2e^{i\theta_2}|11\>-r_3e^{i\theta_3}|10\>.\nn
\eea
One can see by inspection that, applying the gate
$F(\pi/2)$ in standard form is equivalent to applying it to the
corresponding qubits in encoded form - use of the R-I ancilla is
again unnecessary.

Thus, gates of the form $F(\theta)$ are universal for quantum
computing, and, since they can be efficiently implemented by a
gate $G=F(\phi)$ for an appropriately chosen fixed $\phi$, we see
that the gate $G$ is universal for quantum computing within the
circuit model, although it is clearly \emph{not} universal for
performing arbitrary unitary transforms. Such considerations may
be useful for practical implementations, as well as for probing
the more interesting questions about precisely where quantum
computers gain their power and to what extent the standard complex
Hilbert space formulation of quantum mechanics can actually be
argued as necessary as opposed to merely sufficient.

\begin{acknowledgements}
We thank M. Nielsen for pointers to the literature. This research
was supported by the NSA \& ARO under contract No.
DAAG55-98-C-0040.
\end{acknowledgements}

\end{document}